\documentclass[aps,reprint,twocolumn,superscriptaddress,nofootinbib]{revtex4-1}
\usepackage[utf8]{inputenc}
\usepackage{graphicx}
\usepackage{array}

\usepackage{amsmath,commath}
\usepackage{bm}
\usepackage{multirow}
\usepackage[linktocpage=true,colorlinks=true,urlcolor=blue,plainpages=false,pdfpagelabels=false]{hyperref}

\usepackage{ulem}

\newcommand{\e}[1]{\times10^{#1}}

\def\SPbU{Department of Physics, St. Petersburg State University, Universitetskaya Naberezhnaya 7/9, Saint Petersburg 199034, Russia}
\def\PNPI{Petersburg Nuclear Physics Institute named by B. P. Konstantinov of National Research Centre “Kurchatov Institute”, Gatchina, Leningrad District 188300, Russia}

\begin{document}

\title{Spontaneous vacuum decay in low-energy collisions of heavy nuclei beyond the monopole approximation}

\author{R.~V. Popov}
\affiliation{\PNPI}
\affiliation{\SPbU}
\author{V.~M. Shabaev}
\affiliation{\SPbU}
\affiliation{\PNPI}
\author{I.~A. Maltsev}
\affiliation{\SPbU}
\author{D.~A. Telnov}
\affiliation{\SPbU}
\author{N.~K. Dulaev}
\affiliation{\SPbU}
\affiliation{\PNPI}
\author{D.~A. Tumakov}
\affiliation{\SPbU}

\begin{abstract}
    The problem of spontaneous vacuum decay in low-energy collisions of heavy nuclei is considered beyond the scope of the monopole approximation. The time-dependent Dirac equation  is solved in a rotating coordinate system with $z$-axis directed along the internuclear line and the origin placed at the center of mass. The probabilities of electron-positron pair creation and the positron energy spectra are calculated in the approximation neglecting the rotational coupling. The two-center potential is expanded over spherical harmonics and the convergence with respect to the number of terms in this expansion is studied. The results show that taking into account the two-center potential instead of its spherically symmetric part preserves all the signatures of the transition to the supercritical regime that have been found in the framework of the monopole approximation and even enhances some of them.
\end{abstract}
\maketitle

\section{Introduction}

Quantum electrodynamics (QED) in the presence of superstrong electromagnetic fields predicts a number of nonlinear and nonperturbative effects such as light-by-light scattering, vacuum birefringence and production of electron-positron pairs (see, e.g., reviews \cite{ehlotzky2009,rufini2010,dipiazza2012,fedotov2022}). Experimental observation of these effects is complicated by extremely high requirements on the field strength needed for their manifestation. One of the ways to attain such fields relies on ever evolving laser technologies. Although laser facilities in the near future might meet requirements for some of the effects, vacuum pair production is still far from being experimentally accessible. An alternative approach suggests to use heavy nuclei as a source of strong electric field.

In a pioneering work \cite{pom45} it was shown that the $1s$ level of a hydrogen-like ion with an extended nucleus continuously goes down with increasing nuclear charge until at a certain value $Z_{\rm cr}$ it reaches the border of the negative-energy continuum. It raised the question of what happens to a bound state when it joins the positron continuum. In works of Soviet and German physicists~\cite{gerstein1969,pieper1969} it was conjectured that the diving of an initially empty bound state into the negative-energy continuum can result in spontaneous reconstruction of the QED vacuum accompanied with creation of electron-positron pairs (for details see, e.g., Refs.~\cite{popov1970,zel71,mul72,mur76,pop76,mul76,rei77,sof77,migdal1978}). A realistic scenario for observation of this process can be realized in low-energy collision of two heavy nuclei with the total charge exceeding the critical value $Z_1 + Z_2 > Z_{\rm cr}$ \cite{gerstein1969}. When during such collisions the nuclei get sufficiently close to each other, $1s\sigma$ state of the quasimolecule, formed by them, enters the negative-energy continuum as a resonance. As a result, if $1s\sigma$ state was unoccupied, an additional hole enters the lower continuum. Initially localized near the nuclei, this hole can escape to infinity as a free positron, and the initially neutral vacuum becomes charged. This process is known as the spontaneous vacuum decay.

Spontaneous vacuum decay in heavy-ion collisions was a subject of intense theoretical and experimental investigations (see, e.g., reviews \cite{rafelski1978,greiner1985,bosh1986,mullernehler1994,reinhardt2005,rafelski2016} and references therein). The first theoretical calculations of pair creation in the supercritical collisions were carried out in the static approximation, according to which the pair-creation probability is proportional to the time integral of the resonance width $\Gamma(R)$ taken along the nuclear trajectory $ R(t)$~\cite{popov1973,peitz73, popov1979}. Within this approximation, the total probability of spontaneous pair creation, associated with the resonance decay, energy spectra of the emitted positrons as well as their angular distributions were obtained. In Ref.~\cite{popov1979}, a correction for the nonadiabaticity of the tunneling process was also considered. However, the static approach does not take into account the dynamical pair creation induced by the time-dependent potential of the moving nuclei. It turns out that the supercritical resonance has a rather long lifetime, compared to the duration of the supercritical regime $\tau_{\rm cr}$. For example, in collisions of uranium nuclei at the energies near the Coulomb barrier (when the nuclei touch each other) the resonance lifetime is about two orders of magnitude larger than $\tau_{\rm cr}$. This makes the probability of spontaneous pair creation quite small. Moreover, the additional width $\Gamma_{\rm dyn} \sim \hbar / \tau_{\rm cr}$, caused by the uncertainty principle, prevents appearance of narrow resonance structures in the energy distribution of the emitted positrons, predicted in the static approximation. Therefore, in order to verify the possibility to observe the signal from the vacuum decay, one needs to take into account the dynamical pair production.

Both the spontaneous and the dynamical mechanisms were investigated by the Frankfurt group (see, e.g., \cite{smith74, reinhardt1981,muller1988}).  From the obtained results it was eventually concluded that experimental observation of spontaneous vacuum decay is possible only if the colliding nuclei would stick to each other for some time due to nuclear forces~\cite{reinhardt2005,rafelski2016}. However, since no evidence of such sticking have been registered to date, this scenario also does not seem promising.

In view of the upcoming experimental facilities in Germany (GSI/FAIR) \cite{gumberidze2009,lestinsky2016}, China (HIAF) \cite{ma2017}, and Russia (NICA) \cite{terAkopian2015} the interest to this problem was renewed. New investigations concerned both static and dynamic aspects of spontaneous positron emission. The properties of the supercritical resonance were addressed for spherically symmetric \cite{ackad2007a,kuleshov2015,godunov2017,krylov2020} \footnote{Although we acknowledge calculations of supertcritical resonances in Refs.~\cite{kuleshov2015,krylov2020}, we disagree with the conclusion made by the authors about absence of spontaneous pair creation.} and non-symmetric \cite{ackad2007b,marsman2011,maltsev2020} field configurations. The behaviour of the vacuum polarization energy for supercticical Coulomb fields was examined in a series of papers, see, e.g., \cite{grashin2022,krasnov2022} and references therein. Dynamic consideration of pair-creation in heavy-nuclei collisions was targeted in the framework of the monopole approximation \cite{ackad2008,bondarev2015,maltsev2015} and beyond \cite{maltsev2017,maltsev2018,popov2018}. Relativistic semiclassical approach was applied to the vacuum instability problem in Ref.~\cite{voskresensky2021}.

Recently there was proposed a new way to see the signs indicating the transition to the supercritical regime, where spontaneous electron-positron pair creation becomes possible \cite{maltsev2019,popov2020}. The method suggests to consider collisions along trajectories corresponding to different energies but having the same distance of the closest approach, $R_{\rm min}$. As the parameters that define the specific trajectory, it is convenient to use $R_{\rm min}$ and the ratio $\eta = E/E_0$  $\in[1, \infty)$ of the collision energy $E$ to the energy of the head-on collision with the same $R_{\rm min}$. The idea behind this is the opposite dependence of the dynamic and spontaneous contributions to the pair-creation probability on the nuclear velocity, characterized here by the parameter $\eta$. Indeed, it is clear that the contribution of the spontaneous mechanism is determined by the time $\tau_{\rm cr}$ the nuclei spend in the region $R_{\rm min} \le R(t) < R_{\rm cr}$, where $R(t)$ is the internuclear distance and $R_{\rm cr}$ is the distance at which the $1$s$\sigma$ state of the quasimolecule reaches the negative-energy continuum, i.e., $E_{1s\sigma}(R_{\rm cr}) = -m_ec^2$ with $m_e$ being the electron mass. This time monotonically decreases with the increase of collision energy, i.e., $\eta$, and so does the contribution of the spontaneous mechanism. On the contrary, the dynamical pair production should increase with the increase of $\eta$. Therefore, the raise of the pair-creation probability with $\eta \to 1$ is to be attributed to the transition to the supercritical regime and activation of the spontaneous mechanism. More details are to be found in Ref.~\cite{popov2020}.

By employing the aforementioned approach, the detailed investigation of the $\eta$-dependence of the pair-production probabilities and positron energy spectra was carried out in Ref.~\cite{popov2020} and later independently confirmed in Ref.~\cite{reus2022}. The calculations were conducted within the monopole approximation, where only spherically symmetric part of the two-center nuclear potential is taken into account. The evidence of the transition to the supercritical regime have been found in both the pair-creation probabilities and positron spectra. Although it has been shown that the monopole approximation works rather well for description of the pair-creation process \cite{maltsev2017,maltsev2018,popov2018}, it is important to study how consideration of the two-center potential would affect the signs of the transition to the supercritical regime mentioned above. Also, calculations beyond the monopole approximations are necessary to get access to other important aspects of nuclei collisions, e.g., the angular resolved positron spectra. To this end, in this work we performed the calculations taking into account higher-order terms in the decomposition of the nuclear potential over spherical harmonics. The calculations are performed in the coordinate system with $z$-axis directed along the internuclear line and the origin located at the center of mass. The rotational-coupling term that appears in the time-dependent Dirac equation due to the transition to this noninertial reference frame (see, e.g., Refs.~\cite{muller1976}) as well as the magnetic field of the nuclei were not taken into account. As it was shown in Ref.~\cite{betz1976,soff1979,reinhardt1980,soff1988}, the influence of these effects on the total probability and positron energy spectra is negligible. It should be noted, however, that the rotational and magnetic terms can have some impact on the positron angular distributions which are not the subject of study of the present work.

The relativistic units ($\hbar=c=1$) and the Heaviside charge unit ($\alpha=e^2/(4\pi)$, $e < 0$) are used throughout the paper.

\section{Theory \label{sec:theory}}

The calculations are based on the formalism of quantum electrodynamics with unstable vacuum developed in Ref.~\cite{fradkin1991}. The nuclei are treated classically as finite-size particles moving along  the hyperbolic Rutherford trajectories. The vector part of the 4-potential created by the nuclei is neglected.

    The pair-creation probabilities and positron energy spectra can be expressed in terms of one-electron transition amplitudes. To calculate the amplitudes, one has to to solve the time-dependent Dirac equation,
    \begin{align}
        \label{eqn:TDDE_exact}
        i\partial_t \psi_i(\bm{r}, t) &= H(t) \psi_i(\bm{r}, t)
    \end{align}
    with
    \begin{align}
        H(t) &= \bm{\alpha}\cdot\bm{p} + \beta m_e + V(\bm{r}, t),
    \end{align}
    where $\bm{\alpha}$, $\beta$ are the Dirac matrices, the subscript $i$ specifies the initial condition, and $V(\bm{r}, t)$ is the total two-center potential generated by the colliding nuclei,
    \begin{align}
        V(\bm{r}, t) &=  V_{\rm A}\left(|\bm{r}-\bm{R}_{\rm A}(t)|\right) + V_{\rm B}\left(|\bm{r}-\bm{R}_{\rm B}(t)|\right).
    \end{align}
    Here $\bm{R}_{\rm A/B}(t)$ denotes the nuclear coordinates. In our calculations we utilize an expansion of the time-dependent wave function over a finite static basis set $\{u_j(\bm{r})\}_{j=1}^N$:
    \begin{align}
        \label{eqn:td_psi_expansion}
        \psi_i(\bm{r}, t) = \sum_j a_{ji}(t)u_j(\bm{r}).
    \end{align}
    The basis set $\{u_j(\bm{r})\}_{j=1}^N$ consists of a number of subsets $\{u_j^{\kappa}(\bm{r})\}_{j=1}^n$ containing functions of certain angular symmetry described by the angular-momentum--parity quantum number $\kappa$. Functions $u_j^{\kappa}$ are bispinors with radial parts represented by B-splines in accordance with the dual kinetic balance (DKB) approach \cite{shabaev2004}. Each subset of $u_j^{\kappa}$, pertaining to certain $\kappa$, is split into two parts. The first part with $1 \le j \le n/2$ is defined as
    \begin{align}
        \label{eqn:expansion_basis_upper}
        u_j^{\kappa}(\bm{r}) &= \frac{1}{r}\left(\begin{array}{r}
                                                          B_j(r)\Omega_{\kappa\mu}(\bm{n}) \\
                                      \frac{1}{2m_e}\left(\frac{d}{dr}+\frac{\kappa}{r}\right) B_j(r)\Omega_{-\kappa\mu}(\bm{n})\\
                                    \end{array}
                               \right)
    \end{align}
    and the second one with $ n/2 < j \le n$ reads
    \begin{align}
        \label{eqn:expansion_basis_lower}
        u_j^{\kappa}(\bm{r}) &= \frac{1}{r}\left(\begin{array}{r}
                                      \frac{1}{2m_e}\left(\frac{d}{dr}-\frac{\kappa}{r}\right) B_j(r)\Omega_{\kappa\mu}(\bm{n}) \\
                                                          B_j(r)\Omega_{-\kappa\mu}(\bm{n})\\
                                    \end{array}
                               \right).
    \end{align}
    Here $B_j(r)$ is the $j$th B-spline, $\Omega_{\kappa\mu}(\bm{n})$ is the spherical spinor, and $\bm{n}=\bm{r}/r$. This choice of basis functions is highly advantageous in the case of symmetric collisions, where the odd harmonics in the multipole expansion of the two-center potential,
    \begin{align}
        V(\bm{r}, t) = \sum_{L=0}^{\infty}\sum_{M=-L}^L \sum_{\alpha={\rm A,B}}V^{\alpha}_{LM}\left(r, \bm{R}_{\alpha}(t)\right) Y_{LM}(\bm{n}),
    \end{align}
    where
    \begin{align}
        V^{\alpha}_{LM}\left(r, \bm{R}_{\alpha}(t)\right) = \int d\bm{n}\; Y_{LM}^{\ast}(\bm{n})\; V_{\alpha}\left(|\bm{r}-\bm{R}_{\alpha}(t)|\right),
    \end{align}
    cancel out in the center-of-mass frame. Thus, the states with opposite spatial parity become decoupled and can be propagated independently. This, in turn, reduces the size of matrices describing the discretized version of Eq.~\eqref{eqn:TDDE_exact} (see below) by almost a half, which significantly facilitates the computations.

    When using a finite basis set, the initial Eq.~\eqref{eqn:TDDE_exact} is transformed to a system of ordinary differential equations:
    \begin{align}
        \label{eqn:TDDE_matrix_form}
        i S\frac{\partial \bm{a}_{i}(t)}{\partial t} = H(t)\bm{a}_{i}(t),
    \end{align}
    where $\bm{a}_i = \{a_{1i},\ldots,a_{Ni}\}$ denotes the array of expansion coefficients, $S_{jk} = \langle u_j | u_k \rangle$ is the overlap matrix, and $H_{jk}(t) = \langle u_j | H(t) | u_k \rangle$ is the Hamiltonian matrix. The set of equations \eqref{eqn:TDDE_matrix_form} is subsequently solved with the aid of the Crank-Nicolson scheme \cite{crank1947}. This scheme imposes the following relation on the coefficients $\bm{a}_i(t)$ taken at adjacent time steps separated by interval $\Delta t$:
   \begin{align}
       \left[S+\frac{i\Delta t}{2}H(t+\Delta t/2)\right]\bm{a}_{i}(t+\Delta t) =\nonumber\\
       \left[S-\frac{i\Delta t}{2}H(t+\Delta t/2)\right]\bm{a}_{i}(t).
   \end{align}

   To further simplify the calculations we use the coordinate system, whose $z$-axis is tied to the internuclear line and rotates together with it. Meanwhile, the rotational-coupling term -- $\bm{j}\cdot\bm{\omega}$ ($\bm{j}$ is the electronic angular momentum and $\bm{\omega}$ is the angular velocity vector) that appears in the Hamiltonian upon the transformation \cite{muller1976} is neglected. In this coordinate system we can use the eigenfunctions $\varphi_i$ of $H(t_{\rm in})=H(t_{\rm out})$ as the initial and final states. These eigenfunctions are found from the matrix version of the stationary Dirac equation. Using the expansion of $\varphi_i$ similar to Eq.~\eqref{eqn:td_psi_expansion} with the coefficients $c_{k}$, one arrives at the following generalized eigenvalue problem:
   \begin{align}
        \label{eqn:h0_eigen_problem}
       H\bm{c} = \epsilon S\bm{c},
   \end{align}
   where $H_{jk}=\langle u_j | H(t_{\rm in}) | u_k \rangle$ and $\bm{c} = \{c_{1},\ldots,c_{N}\}$. Solving Eq.~\eqref{eqn:h0_eigen_problem} yields a set of eigenvalues $\epsilon_i$ and eigenvectors $\bm{c}_i$ ($i=1,\ldots,N$) which represent a discretized version of the $H(t_{\rm in})$ spectrum. The initial conditions for Eq.~\eqref{eqn:TDDE_matrix_form} are then set as
   \begin{align}
       \bm{a}_i(t_{\rm in}) = \bm{c}_i.
   \end{align}

   The one-electron transition amplitudes attain the form
   \begin{align}
       A_{fi} &= \langle \varphi_f | \psi_i(t_{\rm out})\rangle\nonumber\\ &=\bm{c}_f^{\dagger}S\bm{a}_i(t_{\rm out}).
   \end{align}
   Finally, the mean number of positrons created in the $m$th energy state is \cite{fradkin1991,greiner1985}
   \begin{align}
        \label{eqn:mean_num_of_positrons}
       \overline{n}_{m} = \sum_{\epsilon_j > -1}|A_{mj}|^2.
   \end{align}

    The calculations of positron energy spectra were performed with the modified Stieltjes procedure \cite{langhoff1974,maltsev2018}:
    \begin{align}
    \label{eqn:stieltjes}
     &\frac{dP}{d\varepsilon}\Bigl(\frac{\varepsilon_p+\varepsilon_{p+N_s-1}}{2}\Bigr) \nonumber\\
     &=\frac{1}{\varepsilon_{p+N_s-1}-\varepsilon_{p}}\left(\frac{\overline{n}_{p}+\overline{n}_{p+N_s-1}}{2}+\sum_{i=1}^{N_s-2}\overline{n}_{p+i}\right)\,.
   \end{align}
    Here $N_s$ determines the number of energy eigenvalues involved in the calculation of one point in the spectrum. With $N_s=2$ Eq.~\eqref{eqn:stieltjes} turns into the regular Stieltjes formula. We used $N_s$ equal to a multiple of the number of the utilized $\kappa$ channels.

\section{Results}

Following the method described above, we performed calculations of the pair-creation probabilities and positron energy spectra for collisions of bare nuclei with various charge numbers. The nuclei were treated classically as homogeneously charged spheres of radius $R_{\rm n}=1.2A^{1/3}$ fm, where $A$ is the atomic mass number. Their motion was described by the hyperbolic trajectories. As was demonstrated in Ref.~\cite{maltsev2018}, when the rotation of the internuclear axis is neglected, the dominant contribution to the probability comes from states with angular momentum projections $|\mu|=\frac{1}{2}$. Therefore, only states with $\mu=\frac{1}{2}$ were included into the basis set and the results were doubled. The basis functions \eqref{eqn:expansion_basis_upper}, \eqref{eqn:expansion_basis_lower} were constructed with B-splines of the 9th order generated on the grid of size $R_{\rm box}=68.5$ r.u.  The nodes were distributed polynomially with $r_i=R_{\rm box}\left(i/(N-1)\right)^4$. The initial and final internuclear distance was taken to be $R(t_{\rm in}) = R(t_{\rm out})\equiv R_0=5000$ fm. The number of propagated electron states was reduced by introducing a cutoff energy $\varepsilon_{\rm c}=6$ r.u. Only states with energy $\varepsilon \in (-1,\ \varepsilon_c]$ were taken into account in Eq.~\eqref{eqn:mean_num_of_positrons}, providing the relative inaccuracy of the sum on the level of $10^{-4}$.

\subsection{Pair-creation probabilities}
First, we studied the dependence of the pair-creation probability on the number of the $\kappa$ channels included in the expansion \eqref{eqn:td_psi_expansion} of the time-dependent wave function. For this purpose we considered collisions of bare uranium nuclei at the energy of $6.218$ MeV/u. Table~\ref{tab:prob_kappa_convergence} contains the total pair-creation probability $P_{\rm t}$ and the contributions of the ground ($P_{\rm g}$) and all bound states ($P_{\rm b}$) obtained for several impact parameters in the range from 0 to 30 fm. For comparison the values calculated in Ref.~\cite{maltsev2018} are also presented. The table shows a rather fast convergence of the total probability with respect to the number of the $\kappa$ channels. For example, the basis with $|\kappa|_{\rm max}=3$ already provides a deviation from the converged results of less than $1$\%. Thus, in further calculation only functions with $|\kappa|\le 3$ were included in the basis.
\begin{table*}
  \centering
  \caption{Dependence of the pair creation probability on $|\kappa|_{\rm max}$ for collisions of bare uranium nuclei at the energy of $6.218$ MeV/u. $P_{\rm t}$ is the total pair-creation probability, $P_{\rm g}$ and $P_{\rm b}$ are the contributions of the ground and all bound states, respectively. The entries with $|\kappa|_{\rm max}=1$ correspond to the monopole approximation.}
   \label{tab:prob_kappa_convergence}
  \begin{tabular}{cc*{6}{>{\centering}p{0.11\linewidth}}>{\centering\arraybackslash}p{0.11\textwidth}}
    \hline
    \hline
              & $|\kappa|_{\rm max}$ &         \multicolumn{7}{c}{Impact parameter (fm)}  \\
     \cline{3-9}
            &          &       0      &       5      &      10      &     15       &      20      &     25       &      30      \\
 \hline
            &     $1$  & $1.04\e{-2}$ & $8.80\e{-3}$ & $6.02\e{-3}$ & $3.84\e{-3}$ & $2.41\e{-3}$ & $1.51\e{-3}$ & $9.50\e{-4}$ \\
            &     $3$  & $1.09\e{-2}$ & $9.24\e{-3}$ & $6.41\e{-3}$ & $4.15\e{-3}$ & $2.64\e{-3}$ & $1.68\e{-3}$ & $1.07\e{-3}$ \\
            &     $5$  & $1.11\e{-2}$ & $9.46\e{-3}$ & $6.58\e{-3}$ & $4.27\e{-3}$ & $2.73\e{-3}$ & $1.74\e{-3}$ & $1.11\e{-3}$ \\
$P_{\rm g}$ &     $7$  & $1.10\e{-2}$ & $9.34\e{-3}$ & $6.50\e{-3}$ & $4.23\e{-3}$ & $2.70\e{-3}$ & $1.73\e{-3}$ & $1.11\e{-3}$ \\
            &     $9$  & $1.08\e{-2}$ & $9.24\e{-3}$ & $6.42\e{-3}$ & $4.18\e{-3}$ & $2.67\e{-3}$ & $1.71\e{-3}$ & $1.10\e{-3}$ \\
            &    $11$  & $1.08\e{-2}$ & $9.19\e{-3}$ & $6.39\e{-3}$ & $4.16\e{-3}$ & $2.66\e{-3}$ & $1.70\e{-3}$ & $1.09\e{-3}$ \\
&Ref. \cite{maltsev2018} & $1.09\e{-2}$ & $9.30\e{-3}$ & $6.47\e{-3}$ & $4.21\e{-3}$ & $2.73\e{-3}$ & $1.72\e{-3}$ & $1.11\e{-3}$ \\
\hline
            &     $1$  & $1.25\e{-2}$ & $1.05\e{-2}$ & $7.03\e{-3}$ & $4.39\e{-3}$ & $2.70\e{-3}$ & $1.66\e{-3}$ & $1.03\e{-3}$ \\
            &     $3$  & $1.32\e{-2}$ & $1.12\e{-2}$ & $7.63\e{-3}$ & $4.85\e{-3}$ & $3.03\e{-3}$ & $1.89\e{-3}$ & $1.19\e{-3}$ \\
            &     $5$  & $1.32\e{-2}$ & $1.11\e{-2}$ & $7.62\e{-3}$ & $4.86\e{-3}$ & $3.05\e{-3}$ & $1.91\e{-3}$ & $1.21\e{-3}$ \\
$P_{\rm b}$ &     $7$  & $1.31\e{-2}$ & $1.11\e{-2}$ & $7.59\e{-3}$ & $4.84\e{-3}$ & $3.04\e{-3}$ & $1.91\e{-3}$ & $1.21\e{-3}$\\
            &     $9$  & $1.31\e{-2}$ & $1.11\e{-2}$ & $7.58\e{-3}$ & $4.83\e{-3}$ & $3.03\e{-3}$ & $1.90\e{-3}$ & $1.21\e{-3}$ \\
            &    $11$  & $1.31\e{-2}$ & $1.11\e{-2}$ & $7.58\e{-3}$ & $4.83\e{-3}$ & $3.03\e{-3}$ & $1.90\e{-3}$ & $1.20\e{-3}$ \\
&Ref. \cite{maltsev2018}  & $1.32\e{-2}$ & $1.12\e{-2}$ & $7.64\e{-3}$ & $4.87\e{-3}$ & $3.07\e{-3}$ & $1.93\e{-3}$ & $1.23\e{-3}$ \\
\hline
            &     $1$  & $1.29\e{-2}$ & $1.08\e{-2}$ & $7.26\e{-3}$ & $4.51\e{-3}$ & $2.75\e{-3}$ & $1.69\e{-3}$ & $1.04\e{-3}$ \\
            &     $3$  & $1.36\e{-2}$ & $1.15\e{-2}$ & $7.83\e{-3}$ & $4.95\e{-3}$ & $3.08\e{-3}$ & $1.92\e{-3}$ & $1.20\e{-3}$ \\
            &     $5$  & $1.36\e{-2}$ & $1.15\e{-2}$ & $7.81\e{-3}$ & $4.96\e{-3}$ & $3.10\e{-3}$ & $1.94\e{-3}$ & $1.22\e{-3}$ \\
$P_{\rm t}$ &     $7$  & $1.35\e{-2}$ & $1.14\e{-2}$ & $7.79\e{-3}$ & $4.95\e{-3}$ & $3.09\e{-3}$ & $1.94\e{-3}$ & $1.22\e{-3}$\\
            &     $9$  & $1.35\e{-2}$ & $1.14\e{-2}$ & $7.78\e{-3}$ & $4.94\e{-3}$ & $3.09\e{-3}$ & $1.93\e{-3}$ & $1.22\e{-3}$ \\
            &    $11$  & $1.35\e{-2}$ & $1.14\e{-2}$ & $7.78\e{-3}$ & $4.94\e{-3}$ & $3.09\e{-3}$ & $1.93\e{-3}$ & $1.22\e{-3}$ \\
&Ref. \cite{maltsev2018} & $1.38\e{-2}$ & $1.16\e{-2}$ & $8.01\e{-3}$ & $5.15\e{-3}$ & $3.46\e{-3}$ & $2.14\e{-3}$ & $1.42\e{-3}$ \\
    \hline
    \hline
  \end{tabular}
\end{table*}

\begin{figure}
    \centering
    \includegraphics[width=\columnwidth]{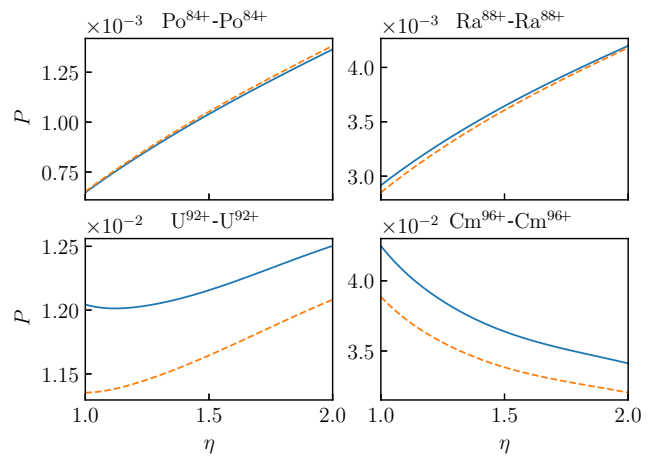}
    \caption{Total pair-creation probability as a function of $\eta$ with $R_{\rm min}=17.5$ fm. Solid blue lines depict results obtained with $|\kappa|_{\rm max}=3$, dashed orange curves correspond to the monopole-approximation results.}
    \label{fig:p_vs_eta_varZ}
\end{figure}

Henceforth we consider the total pair-creation probability and denote it with $P$ omitting the subscript. It was shown in Refs.~\cite{maltsev2019,popov2020} that in the scope of the monopole approximation the pair-creation probability as a function of $\eta$ increases as $\eta\to 1$, when $R_{\rm min}$ and $Z_{\rm t}=Z_1+Z_2$ enter deeply enough into the supercritical domain of collision parameters. This increase can serve as an indication of the transition to the supercritical regime. In this work we studied how the dependence of the probability $P$ on $\eta$ changes when higher-order terms in the potential decomposition are brought into consideration. For $R_{\rm min}=17.5$ fm, the results obtained in the basis with $|\kappa|_{\rm max}=3$ for symmetric collisions of bare nuclei with subcritical ($Z=84$) and supercritical ($Z=88,\ 92,\ 96$) charge numbers are displayed in Fig.~\ref{fig:p_vs_eta_varZ} in comparison with the monopole-approximation results. The comparison shows that the effects associated with higher-order terms somewhat enhance the manifestation of the increase of $P$ as $\eta \to 1$ for supercritical charge numbers.  For instance, in the case of the U$^{92+}$-U$^{92+}$ collisions, the probability obtained with $|\kappa|_{\rm max}=3$ exhibits a shallow minimum near $\eta=1$, which is absent in the monopole approximation.

\begin{figure}
    \centering
    \includegraphics[width=\columnwidth]{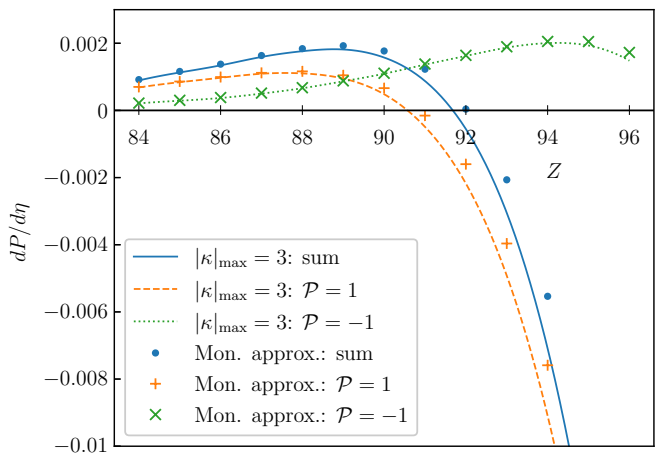}
    \caption{Derivative of the pair-creation probability $dP/d\eta$ at $\eta=1$ as a function of $Z=Z_1=Z_2$.}
    \label{fig:dPdEta_vs_Z}
\end{figure}

The influence of the nonmonopole terms becomes more apparent when considering the derivative of the pair-creation probability with respect to the parameter $\eta$, $dP/d\eta$, at $\eta=1$. Figure \ref{fig:dPdEta_vs_Z} represents the contributions of odd $(\mathcal{P}=-1)$ and even $(\mathcal{P}=1)$ states to $\eval[1]{dP/d\eta}_{\eta=1}$ as functions of $Z$. As can be seen from Fig.~\ref{fig:dPdEta_vs_Z}, the deviation from the monopole results is hardly visible until the corresponding channel becomes supercritical, which happens at $Z\approx 87.3$ for $\mathcal{P}=1$ and $Z\approx 94.8$ for $\mathcal{P}=-1$. In the supercritical region the values of $dP/d\eta$ obtained with $|\kappa|_{\rm max}=3$ lie lower than the monopole ones. This behavior of $dP/d\eta$ aligns with the findings of Refs.~\cite{marsman2011,maltsev2020}, where the supercritical-resonance parameters were examined beyond the monopole approximation. According to Refs.~\cite{marsman2011,maltsev2020}, inclusion of higher-order terms in the potential decomposition results in about $20\%$ increase in the resonance width of U$_2^{183+}$ quasimolecule at the internuclear distance of $16$ fm. Furthermore, this increase in width turns out to be larger for larger internuclear separations. Note that supercritical resonance width is exclusively due to the spontaneous pair creation while in collisions of heavy nuclei both spontaneous and dynamic mechanisms contribute to the total pair-creation probability. As seen in Table~\ref{tab:prob_kappa_convergence}, the overall increase in the pair-creation probability for head-on collisions of uranium nuclei at the energy of $6.218$ MeV/u (which corresponds to the internuclear distance of 16.47 fm) amounts to approximately $5\%$. This may indicate that the relative contribution of the spontaneous mechanism to the total pair production became larger, although the electron-positron pairs are predominately created by the dynamic mechanism. As a result one may observe an enhancement of the signal indicating the transition to the supercritical regime found in $dP/d\eta$, namely the sign change from positive to negative. Another factor that can play a role is the extended duration of the supercritical regime, $\tau_{\rm cr}$, due to the increase in the critical internuclear distance $R_{\rm cr}$.

\subsection{Positron spectra}

\begin{figure}
    \centering
    \includegraphics[width=\columnwidth]{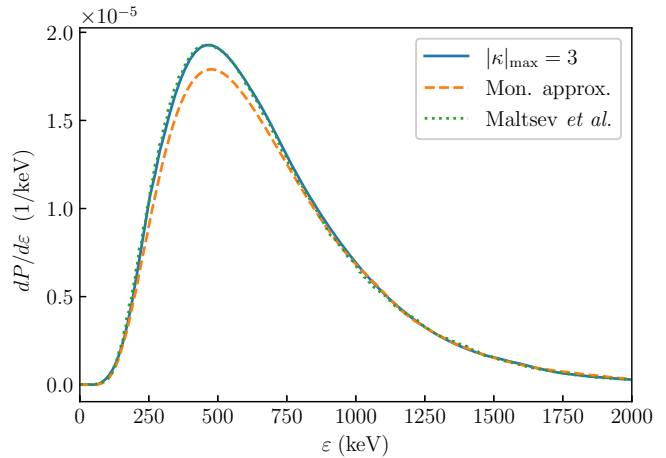}
    \caption{Energy spectra of the positrons emitted in head-on U$^{92+}$-U$^{92+}$ collisions with energy $E=6.218$ MeV/u. Maltsev {\it et al.} refers to \cite{maltsev2018}.}
    \label{fig:z92_pos_spectra_2c}
\end{figure}

Another signature of the transition to the supercritical regime found in Ref.~\cite{popov2020} concerns the $\eta$-dependence of the maximum of the positron energy spectra obtained in collisions with fixed $R_{\rm min}$. It was shown in Ref. \cite{popov2020} in the monopole approximation that in the case of subcritical collisions the spectra corresponding to larger $\eta$ possess higher peak values, whereas for supercritical collisions the dependence is inverted and peak values decrease with increasing $\eta$. In this work we examined whether this behavior remains valid beyond the monopole approximation. At first, we regarded collisions of bare uranium nuclei at the energy of $6.218$ MeV/u. The positron spectra calculated for the head-on collision in the framework of the monopole approximation and beyond it are depicted in Fig.~\ref{fig:z92_pos_spectra_2c}. The spectrum obtained in the basis with $|\kappa|_{\rm max}=3$ is in perfect agreement with the one given in Ref.~\cite{maltsev2018}. The inclusion of higher-order harmonics in the calculations leads to the raise of the spectrum near the peak leaving the tail almost unchanged.

\begin{figure*}
    \centering
    \includegraphics[width=\textwidth]{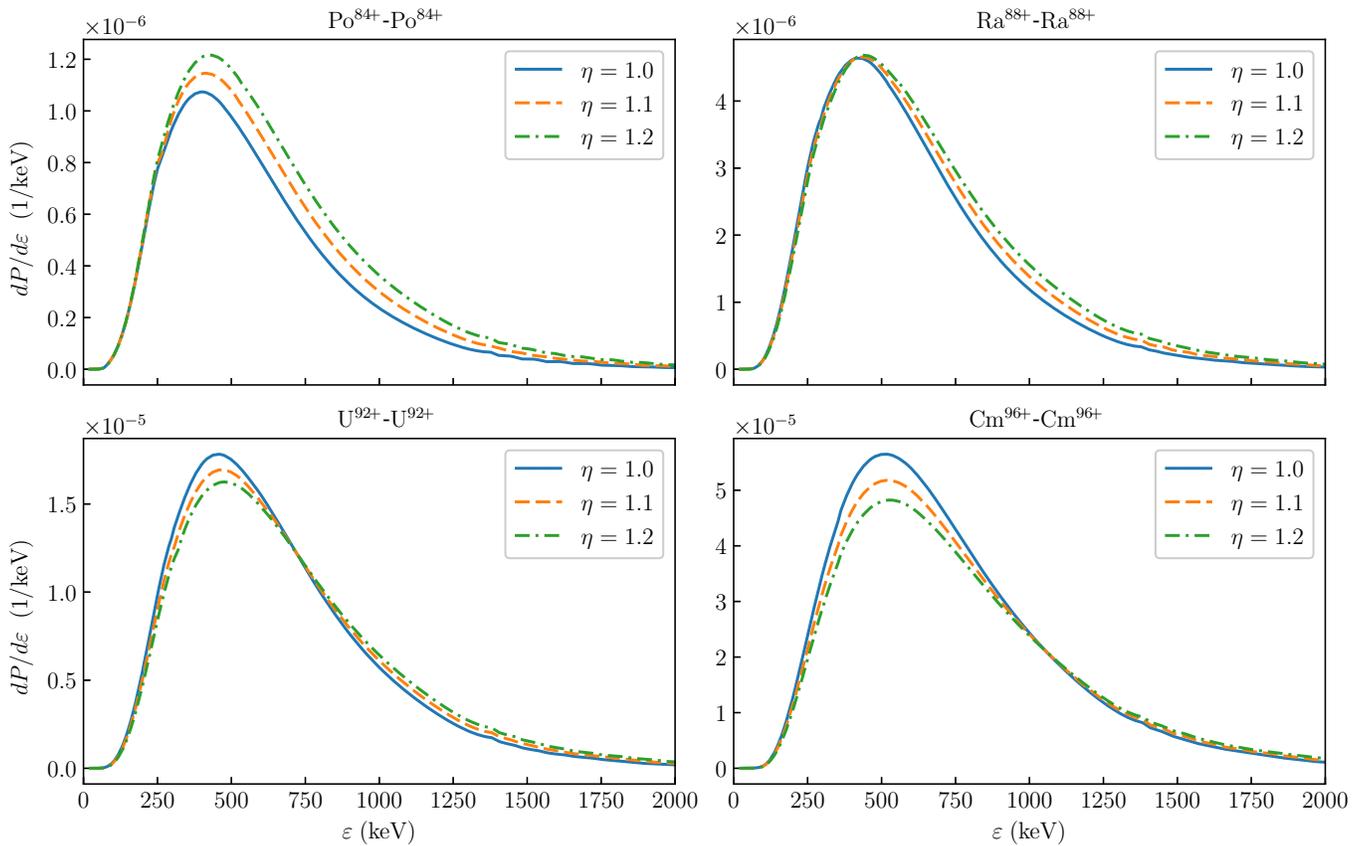}
    \caption{Positron spectra for the symmetric collisions with $Z=Z_1=Z_2=84$--$96$ at $R_{\rm min}=17.5$ fm and $\eta=E/E_0=1,1.1,1.2$.}
    \label{fig:pos_spectra_vs_eta_varZ_2c}
\end{figure*}

After that, we studied the dependence of the positron spectra on $\eta$ for symmetric collisions with a fixed distance of the closest approach, $R_{\rm min}$. In Fig.~\ref{fig:pos_spectra_vs_eta_varZ_2c} we present the spectra obtained for collisions of nuclei with charge numbers $Z=84,\ 88,\ 92,\ 96$, $R_{\rm min}=17.5$ fm, and $\eta=1,\ 1.1,\ 1.2$. The results show that once the total charge number $2Z$ exceeds the critical value, the order of the curves near the peak gets reversed. In full accordance with Ref.~\cite{popov2020}, the subcritical collisions yield higher peak values of the positron spectrum for larger $\eta$, while in the case of the supercritical collisions the opposite relation between the peak hight and $\eta$ is established. The same behavior of the spectra with respect to $\eta$ is found when the supercritical domain of the collision parameters is approached from a different direction, namely when $Z$ is fixed and $R_{\rm min}$ is decreasing.

\section{Conclusion}
We have examined the possibility to access QED in supercritical Coulomb field that can be attained in low-energy collisions of heavy nuclei. The procedure for solving the time-dependent Dirac equation, previously restricted to the monopole approximation, was extended to take into account higher-order terms in the decomposition of the two-center nuclear potential over spherical harmonics. Using this modified procedure, we performed calculations of the pair-creation probabilities and positron energy spectra for collisions of bare nuclei. The results obtained for collisions with a fixed distance of the closest approach exhibit the same signatures of the transition to the supercritical regime as in the monopole approximation \cite{maltsev2019,popov2020}. Inclusion of nonmonopole terms into consideration enhances the manifestation of the signatures found in the behavior of the pair-creation probability as a function of the parameter $\eta=E/E_0$ near $\eta=1$.

\section*{Acknowledgments}
The development of the calculation method, the calculations of the total pair-production probabilities and positron energy spectra were supported by the Russian Science Foundation (Grant No. 22-62-00004). The results for the bound-free production probability were independently verified using a different approach by I.~A. Maltsev supported by the Foundation for the Advancement of Theoretical Physics and Mathematics ``BASIS''.

\end{document}